\documentclass[aps,floatfix,amsmath,nofootinbib,amssymb,superscriptaddress]{revtex4-2}

\usepackage{booktabs} 

\usepackage{overpic}
\usepackage{amssymb}
\usepackage{indentfirst}
\usepackage{feynmf}   
\usepackage{slashed}  
\usepackage{cases}
\usepackage{color}
\usepackage{multirow}
\usepackage{epstopdf}
\usepackage{graphicx,color,bm}
\usepackage{epstopdf}

\usepackage[colorlinks,
            citecolor=green,
            anchorcolor=red,
            menucolor=red,
            linkcolor=red,
            filecolor=red,
            runcolor=red,
            urlcolor=blue,
            frenchlinks=red]{hyperref}

\begin{document}

\title{Accessing gluon GTMD $F^g_{1,4}$ via the $\langle\sin(2\phi)\rangle$ azimuthal asymmetry of exclusive $\pi^0$ production in $ep$ collisions}

\author{Chentao Tan}\affiliation{School of Physics, Southeast University, Nanjing
211189, China}

\author{Zhun Lu}
\email{zhunlu@seu.edu.cn}
\affiliation{School of Physics, Southeast University, Nanjing 211189, China}

\begin{abstract}
The longitudinal single-target spin asymmetry in exclusive $\pi^0$ production in $ep$ collisions is a sensitive probe of the imaginary part of the gluon generalized transverse momentum dependent distribution $F_{1,4}^g$. It appears as a characteristic $\sin(2\phi)$ azimuthal correlation between the transverse momenta of the scattered electron and the recoil proton, generated by Coulomb–nuclear interference; consequently, the Primakoff process should be included. We compute the relevant gluon distributions in a light-front spectator model of the proton that explicitly incorporates gluonic degrees of freedom. This work presents the first model calculation of the imaginary part of $F_{1,4}^g$ and delivers predictions for the resulting asymmetries in kinematics relevant to the planned Electron–-Ion Colliders (EIC and EicC), providing theoretical predictions for upcoming measurements.
\end{abstract}

\maketitle

\section{Introduction}

Understanding the spin structure of hadrons in terms of QCD degrees of freedom—quarks and gluons—remains a central challenge in hadronic physics. 
This field of research was triggered by the so-called ``proton spin crisis"~\cite{EuropeanMuon:1987isl}, which revealed that the spin of the proton cannot be fully accounted for by the spin of its constituent quarks alone. Over the past three decades, substantial progress has been made~\cite{Jaffe:1989jz,Ji:1996ek,Leader:2013jra,Liu:2015xha,Ji:2020ena,Leader:2021pqf}, yet a complete, quantitative decomposition of the proton spin is still an active area of research. 
Two commonly used decompositions are the kinetic (Ji)~\cite{Ji:1996ek} and canonical (Jaffe–Manohar) ~\cite{Jaffe:1989jz} formulations,
The kinetic decomposition yields a gauge-invariant, frame-independent sum rule, where the total angular momentum of the proton is split into the spin and kinetic orbital angular momentum (OAM) of quarks and gluons. The canonical decomposition, by contrast, is defined in the infinite momentum frame and relates the proton spin to the spin and canonical OAM of partons.

For the quark sector, the spin contribution (approximately 20–30\% of the proton spin) is relatively well constrained by the first Mellin moment of the quark helicity parton distribution function (PDF) \cite{EuropeanMuon:1989yki,deFlorian:2009vb,Nocera:2014gqa,Ethier:2017zbq}. The gluon spin contribution, encoded in the gluon helicity PDF, however, exhibits large uncertainties—particularly in the small-$x$ region \cite{Nocera:2014gqa,STAR:2014wox,deFlorian:2014yva,STAR:2021mqa}—and its precise measurement is a primary scientific objective of the planned EIC and EicC \cite{Boer:2011fh,AbdulKhalek:2021gbh,Aschenauer:2014twa}. 
Compared to parton spin, the OAM of both quarks and gluons is even less well understood, from both theoretical and experimental perspectives. Kinetic OAM can be extracted by subtracting parton spin contributions from the total angular momentum measured in hard exclusive processes~\cite{Ji:1996ek,Ji:1996nm,Collins:1996fb}. Canonical OAM, however, is given by the $\bm{k}_\perp$-moment of the generalized transverse momentum distributions (GTMDs) in the forward limit~\cite{Lorce:2011kd,Hatta:2011ku,Lorce:2011ni}, which holds beyond the tree level up to some power corrections~\cite{Ebert:2022cku,Bertone:2022awq,Echevarria:2022ztg}.

GTMDs extend both collinear PDFs and transverse momentum dependent (TMD) PDFs, encapsulating off-forward momentum transfer and transverse parton momentum, and thus provide a complete description of parton structure in the nucleon \cite{Meissner:2008ay,Meissner:2009ww}. But direct experimental access of GTMDs remains challenging  due to the lack of direct observables. For quarks, the exclusive double Drell-Yan process is the only known channel sensitive to the quark GTMD $F^q_{1,4}$ in the Efremov-Radyushkin-Brodsky-Lepage (ERBL) region \cite{Bhattacharya:2017bvs}, while exclusive $\pi^0$ production in $ep$ collisions has recently been proposed as a probe of $F^q_{1,4}$ in the Dokshitzer-Gribov-Lipatov-Altarelli-Parisi (DGLAP) region~\cite{Bhattacharya:2023hbq}. 
For gluons, theoretical efforts have only recently turned to hard exclusive processes sensitive to the gluon GTMD $F^g_{1,4}$—the key object for canonical gluon OAM—including exclusive dijet production \cite{Ji:2016jgn,Hatta:2016aoc,Bhattacharya:2022vvo}, exclusive double Drell-Yan/quarkonium production in hadron collisions \cite{Bhattacharya:2018lgm,Boussarie:2018zwg}, and exclusive $\pi^0$ production in $ep$ collisions with a longitudinally polarized proton target \cite{Bhattacharya:2023yvo}.

In this work, we focus on the longitudinal single target-spin asymmetry in exclusive $\pi^0$ production ($ep\rightarrow e^\prime p^\prime\pi^0$) with an unpolarized electron beam and a longitudinally polarized proton target, an observable proposed in~\cite{Bhattacharya:2023yvo} that is sensitive to the imaginary part of $F^g_{1,4}$. This asymmetry is characterized by a $\sin(2\phi)$ azimuthal correlation between the transverse momenta of the scattered electron and recoil proton, which arises from the \textit{Coulomb-nuclear interference} between two distinct processes for exclusive $\pi^0$ production: the two-gluon exchange process parameterized by $F^g_{1,4}$ and the quasireal photon exchange Primakoff process ($\gamma^\ast\gamma\rightarrow\pi^0$)~\cite{Primakoff:1951iae,Liping:2014wbp,Kaskulov:2011ab,Lepage:1980fj,
Khodjamirian:1997tk,Jia:2022oyl}. 
Therefore, the $\sin(2\phi)$ azimuthal angular correlation actually arises from the interference amplitudes of the above two processes.

To estimate this observable we employ a light-front spectator model that includes explicit gluonic degrees of freedom~\cite{Lu:2016vqu}. Spectator models with gluon components have been used to study the leading twist gluon TMDs~\cite{Lu:2016vqu,Kaur:2019kpe,Bacchetta:2020vty,Bacchetta:2024fci,Chakrabarti:2023djs}, generalized parton distributions (GPDs)~\cite{Tan:2023kbl,Chakrabarti:2024hwx}, Wigner distributions, and GTMDs at zero skewness~\cite{Tan:2023vvi}. A realistic choice of light-front wave functions (LFWFs) is essential for reliable predictions~\cite{Brodsky:2000ii}. 
We model the nucleon–gluon–spectator vertex with a dipole form factor to encode nonperturbative color dynamics~\cite{Bacchetta:2020vty,Bacchetta:2024fci} and fix model parameters by fitting the unpolarized gluon PDF to the NNPDF3.1 at an initial scale $\mu_0=2\ \mathrm{GeV}$~\cite{NNPDF:2017mvq}. The model has previously been benchmarked against experimental constraints and theoretical requirements~\cite{DAlesio:2015fwo,Anselmino:2008sga,COMPASS:2017ezz,Bacchetta:1999kz,Mulders:2000sh}.
The reliability of this model relies on the consistency between model results and experimental data~\cite{DAlesio:2015fwo,Anselmino:2008sga,COMPASS:2017ezz} or theoretical constraints~\cite{Bacchetta:1999kz,Mulders:2000sh}.

The paper is organized as follows. In Sec.~\ref{Sec:2}, we introduce the kinematics of exclusive $\pi^0$ production in $ep$ collisions and the parameterizations of relevant gluon TMDs and GTMDs. In Sec.~\ref{Sec:3} we derive expressions for the unpolarized and helicity PDFs, the gluon Sivers function, and the GTMD $F^g_{1,4}$ of gluons using the overlap representation of the LFWFs in the light-front spectator model. In Sec.~\ref{Sec:4} we present and discuss the numerical results of the mentioned distribution functions and the azimuthal asymmetries for the EIC and EicC kinematics. We summarize our findings in Sec.~\ref{Sec:5}.

\section{The GTMD $F_{1,4}^g$ and the Sivers function of gluons in exclusive $\pi^0$ production}\label{Sec:2}

\begin{figure}
	\centering
	\includegraphics[width=0.45\columnwidth]{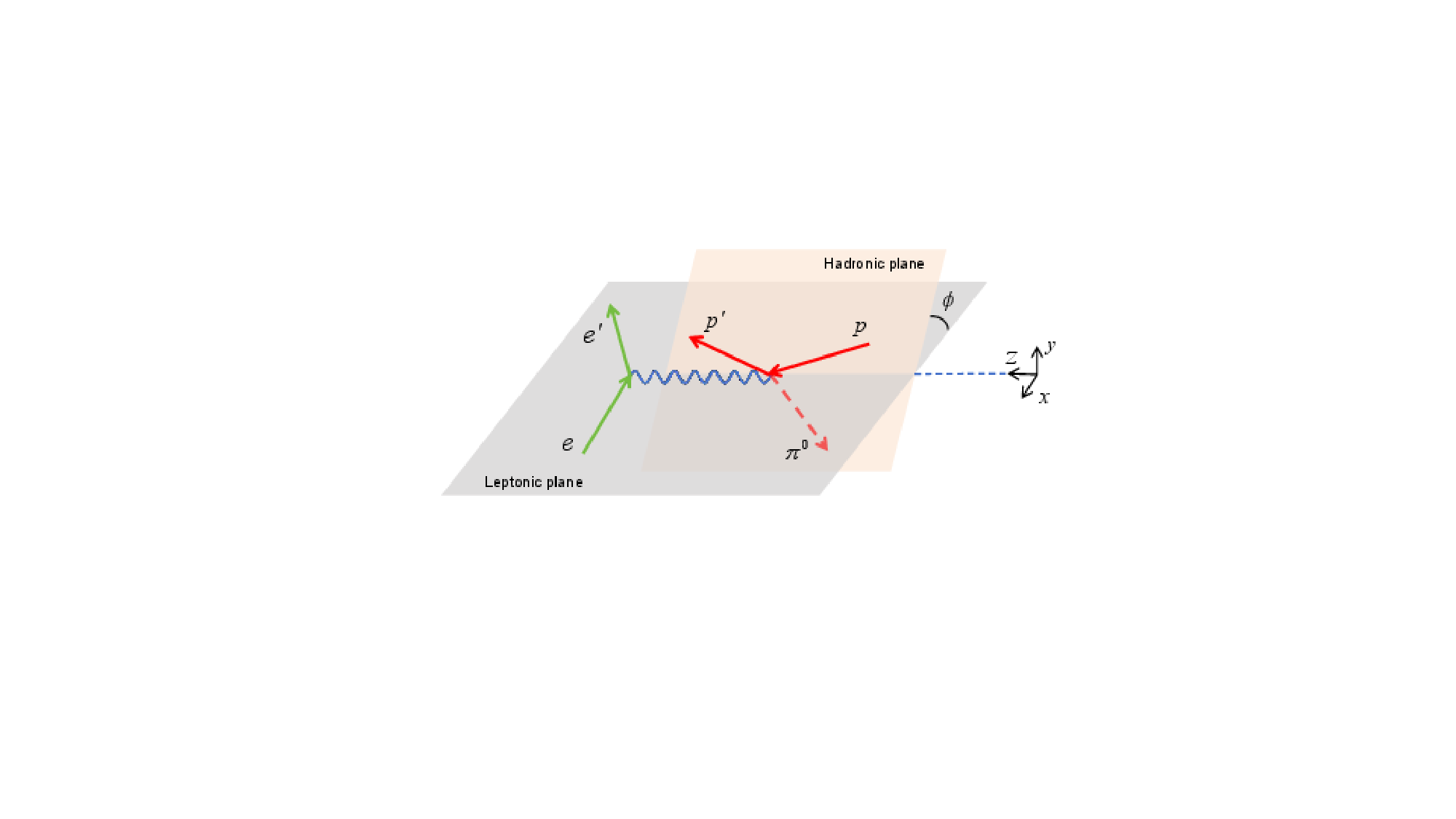}
	\caption{Kinematics of the exclusive $\pi^0$ production in electron-proton collisions.}
	\label{fig:pi0}
\end{figure}

First, we introduce the kinematics of exclusive $\pi^0$ production in electron-proton collisions as shown in Fig.~\ref{fig:pi0}~\cite{Bhattacharya:2023yvo}:
\begin{align}
	e(l)+p(p,\lambda)\to \pi^0(l_\pi)+e(l^\prime)+p(p^{\prime},\lambda^{\prime}),
\end{align}
where $l$ ($l^\prime$) denotes the momentum of the incoming (outgoing) electron, $p$ ($p^\prime$) and $\lambda$ ($\lambda^\prime$) are the momentum and helicity of the incoming (outgoing) proton, respectively, with the proton mass satisfying $p^2=p^{\prime2}=M^2$. The momentum of the produced $\pi^0$ is $l_\pi$, and its mass is neglected for simplicity ($l_\pi^2=0$). We define the following four-momentum combinations for the process: the average proton momentum $P=(p+p^\prime)/2$, the momentum transfer to the proton $\Delta=p^\prime-p$ with $t=\Delta^2=-(4\xi^2M^2+\bm{\Delta}^2_\perp)/(1-\xi^2)$, and the skewness variable $\xi=(p^{+}-p^{\prime+})/(p^{+}+p^{\prime+})=-\Delta^+/2P^+=x_B/(2-x_B)$, where $x_B=Q^2/2p\cdot q$ is the Bjorken variable and $Q^2=-q^2=-(l-l^\prime)^2$ is the photon virtuality. The $ep$ center-of-mass energy squared is $s_{ep}=(p+l)^2$, which defines the momentum fraction $y=p\cdot q/p\cdot l\approx Q^2/(x_B s_{ep})$. The symbol ``$+$'' denotes the light-front plus component, and we work in the symmetric frame with $\bm{P}_\perp=\bm{0}_\perp$.

For a longitudinally polarized proton target, exclusive $\pi^0$ production is dominated by the two-valence-quark exchange process in the region $\xi\sim0.1$ \cite{Bhattacharya:2023hbq}. In this work, we focus on the forward region $t\approx 0$, where the valence quark contribution is negligible and the process is dominated by the two-gluon exchange effect described by $F^g_{1,4}$, and the Primakoff process. At fixed light-front time $z^+=0$, the off-forward gluon-gluon correlator for a spin-$1/2$ target—parameterized by gluon GTMDs—is defined as~\cite{Bhattacharya:2018lgm}
\begin{align}
	W^{[ij]}_{\lambda^\prime\lambda}(P,\Delta,x,\bm{k}_\perp)=\frac{1}{P^+}\int \frac{dz^-d^2\bm{z}_\perp}{(2\pi)^3}e^{ik\cdot z} \bigg\langle p^\prime,\lambda^\prime \bigg| F^{+j}_a\bigg(-\frac{z}{2}\bigg) \mathcal{W}_{ab}\bigg(-\frac{z}{2},\frac{z}{2}\bigg) F^{+i}_b\bigg(\frac{z}{2}\bigg) \bigg| p,\lambda \bigg\rangle \bigg|_{z^+=0},
	\label{eq:Wij}
\end{align}
where $x=k^+/P^+$ and $\bm{k}_\perp$ are the average longitudinal momentum fraction and transverse momentum of the struck gluon, $F_a^{\mu\nu}$ is the QCD field strength tensor (with color index $a$), and $i,j$ are transverse indices. The gauge link $\mathcal{W}_{ab}$ ensures the color gauge invariance of the bilocal gluon operator. At leading twist, there are sixteen independent gluon GTMDs accounting for all parton and nucleon polarization combinations \cite{Lorce:2013pza}; for unpolarized gluons, the correlator simplifies to \cite{Bhattacharya:2018lgm}
\begin{align}
	W_{\lambda^\prime\lambda}=&\delta_\perp^{ij} W^{[ij]}_{\lambda^\prime\lambda} \nonumber\\
	=&\frac{1}{2M}\bar{u}(p^\prime,\lambda^\prime)\bigg[F^g_{1,1}+\frac{i\sigma^{i+}k_\perp^i}{P^+}F^g_{1,2} +\frac{i\sigma^{i+}\Delta_\perp^i}{P^+}F^g_{1,3}+\frac{i\sigma^{ij}k_\perp^i\Delta^j_\perp}{M^2}F^g_{1,4}\bigg]u(p,\lambda) \nonumber\\
	=&\frac{1}{M\sqrt{1-\xi^2}} \bigg\{\bigg[M\delta_{\lambda^\prime,\lambda}-\frac{1}{2}(\lambda\Delta_\perp^1+i\Delta_\perp^2)\delta_{-\lambda^\prime,\lambda}\bigg]F^g_{1,1} +(1-\xi^2)(\lambda k_\perp^1+ik_\perp^2)\delta_{-\lambda^\prime,\lambda}F^g_{1,2} \nonumber \\
	&+(1-\xi^2)(\lambda\Delta_\perp^1+i\Delta_\perp^2)\delta_{-\lambda^\prime,\lambda}F^g_{1,3} +\frac{i\epsilon_\perp^{ij}k_\perp^i\Delta_\perp^j}{M^2} \bigg[\lambda M \delta_{\lambda^\prime,\lambda}-\frac{\xi}{2}(\Delta_\perp^1+i\lambda\Delta_\perp^2)\delta_{-\lambda^\prime,\lambda}\bigg]F^g_{1,4} \bigg\},
	\label{eq:W}
\end{align}	
where $u$ ($\bar{u}$) is the light-front spinor of the incoming (outgoing) proton, $\delta_\perp^{ij}=-g_\perp^{ij}$ with $g^{\mu\nu}$ denoting the metric tensor, and $\epsilon_\perp^{ij}=\epsilon^{-+ij}$ with $\epsilon^{0123}=1$. Gluon GTMDs are a complex-valued functions~\cite{Meissner:2008ay,Meissner:2009ww} of the variables $(x,\xi,\bm{k}_\perp^2,\bm{\Delta}_\perp^2,\bm{k}_\perp\cdot\bm{\Delta}_\perp)$. These GTMDs have been studied in various models~\cite{Tan:2023vvi,Mukherjee:2015aja,Kanazawa:2014nha,Tan:2024dmz}. Of particular interest is the GTMD  $F_{1,4}^g$, which not only defines the canonical gluon OAM~\cite{Hatta:2011ku}, but also quantifies the correlation between the spin of the nucleon and the OAM of the gluon~\cite{Lorce:2011kd,Kanazawa:2014nha}.

For an unpolarized proton target, the cross section of exclusive $\pi^0$ production in the forward-region was once thought to vanish~\cite{Czyzewski:1996bv,Dumitru:2019qec,Engel:1997cga,Harland-Lang:2018ytk}, but Ref.~\cite{Boussarie:2019vmk} demonstrated that at $t\approx0$ this process is a direct probe of the dipole-type gluon Sivers function—a T-odd TMD—via its connection to the spin-dependent Odderon~\cite{Boer:2015pni,Zhou:2013gsa}, which means that the gluon Sivers function (or the spin-dependent Odderon) gives the leading contribution in this case. The gluon Sivers function is inherently linked to the choice of gauge link in the parton correlator, a crucial ingredient for the definition of gluon TMDs. The gauge-link-dependent gluon-gluon correlator is given by ~\cite{Mulders:2000sh,Bomhof:2007xt,Bomhof:2006dp,Buffing:2013kca}:
\begin{align}
	\Phi^{\mu\nu[U,U^\prime]}(x,\bm{k}_\perp,S)=\frac{1}{xP^+}\int\frac{dz^-d^2\bm{z}_\perp}{(2\pi)^3}e^{ik\cdot z}\bigg\langle P,S\bigg|2\text{Tr}\bigg[F^{+\nu}\bigg(-\frac{z}{2}\bigg)\mathcal{U}_{[-\frac{z}{2},\frac{z}{2}]}F^{+\mu}\bigg(\frac{z}{2}\bigg)\mathcal{U}^\prime_{[\frac{z}{2},-\frac{z}{2}]}\bigg]\bigg|P,S\bigg\rangle\bigg|_{z^+=0},
	\label{eq:phi}
\end{align}	
where $\text{Tr}$ denotes a trace over color-triplet indices. The two staple-like gauge links $\mathcal{U}$ and $\mathcal{U}^\prime$ arise from the color-triplet representation of the field operators \cite{Bomhof:2006dp}: the future-pointing gauge link describes final-state interactions between the spectator and outgoing gluon, while the past-pointing gauge link describes initial-state interactions. For $\mathcal{U}^\prime=\mathcal{U}^\dagger$, the correlator in Eq.~(\ref{eq:phi}) reduces to expression in the adjoint representation of the $\text{SU}(3)$ color group (Eq.~\ref{eq:Wij}), written as $\langle F_a\mathcal{U}_{ab}F_b\rangle$.

Closed loops composed of the gauge links $\mathcal{U}$ and $\mathcal{U}^\prime$ are usually denoted by their directions~\cite{Bomhof:2006dp,Bacchetta:2024fci}: $[+,+]$ for future-pointing loops (e.g., semi-inclusive deep-inelastic scattering (SIDIS) dijet production~\cite{Pisano:2013cya}) and $[-,-]$ for past-pointing loops (e.g., Higgs production via gluon fusion~\cite{Boer:2013fca,Echevarria:2015uaa}). The TMDs from these gauge-link loops are called as Weizs$\ddot{\text{a}}$cker-Williams ($f$-type) gluon TMDs, with T-odd components involving the antisymmetric color structure constants $f^{abc}$ \cite{Dominguez:2010xd,Dominguez:2011wm}. Loops with mixed forward/past-pointing gauge links ($[+,-]$ and $[-,+]$)—sensitive to the direction of color flow (e.g., photon-jet production in SIDIS \cite{Bomhof:2006dp,Bacchetta:2007sz,Buffing:2018ggv})—yield dipole ($d$-type) gluon TMDs, with T-odd components involving the symmetric color structure constants $d^{abc}$ \cite{Dominguez:2010xd,Dominguez:2011wm}. T-even gluon TMDs are invariant under gauge-link direction, i.e.,  $[+,+]=[-,-]$ for $f$-type TMDs and $[+,-]=[-,+]$ for $d$-type TMDs, while T-odd TMDs change sign.

In general, $f$-type and $d$-type gluon TMDs are independent, encoding distinct physical information and requiring extraction from different processes~\cite{Boer:2015vso}. However, Ref.~\cite{Bacchetta:2024fci} found a universal ratio $[+,-]/[+,+]=5/9$ for two spectator models, a result we adopt in this work. We first calculate the $f$-type Sivers function from the correlator $\Phi^{ij[+,+]}$, then obtain the $d$-type Sivers function via this ratio. The leading-twist $f$-type gluon TMDs are defined from $\Phi^{ij[+,+]}$ as~\cite{Meissner:2007rx}
\begin{align}
	\Phi^{[+,+]}(x,\bm{k}_\perp,S)=&\delta_\perp^{ij}\Phi^{ij[+,+]}(x,\bm{k}_\perp,S)\nonumber\\
	=&f_1^{g(f)}(x,\bm{k}_\perp^2)-\frac{\epsilon_\perp^{ij}k_\perp^iS_\perp^j}{M}f_{1T}^{\perp g(f)}(x,\bm{k}_\perp^2),\\
	\tilde{\Phi}^{[+,+]}(x,\bm{k}_\perp,S)=&i\epsilon_\perp^{ij}\Phi^{ij[+,+]}(x,\bm{k}_\perp,S)\nonumber\\
	=&\lambda g_{1L}^{g(f)}(x,\bm{k}_\perp^2)+\frac{\bm{k}_\perp\cdot\bm{S}_\perp}{M}g_{1T}^{g(f)}(x,\bm{k}_\perp^2).
\end{align}	
We omit the superscript ``$f$'' in the following unless explicitly stated. 
Finally, we note that exclusive $\pi^0$ production in $ep$ collisions with a transversely polarized proton target is sensitive to leading-twist helicity-flip quark GPDs \cite{Frankfurt:1999fp}, a topic beyond the scope of this work.

\section{Gluon distribution functions in the overlap representation}\label{Sec:3}

To investigate the gluon distribution functions in the proton, we adopt the light-front spectator model from Ref.~\cite{Lu:2016vqu}. In the model, the struck gluon is radiated from the proton, making the minimal Fock state of the proton $|qqqg\rangle$. To simplify the four-body system, we treat the proton as a two-particle composite state of a spin-1 active gluon $g$ and a spin-1/2 spectator $X$ (comprising the three valence quarks):
\begin{align}
	|p,S \rangle \rightarrow |g^{s_g}X^{s_X}(uud)\rangle ,
\end{align}
where $s_g$ and $s_X$ are the spins of the gluon and spectator, respectively. In the naive quark model, $s_X$ can be $1/2$ or $3/2$, and higher values are possible if inter-quark OAM is included. However, the spectator—an isolated, thermodynamically balanced quantum system—is more likely to occupy its ground state, so we only consider $s_X=1/2$ and neglect higher-spin contributions.

For the proton states with the helicities $J_z=\pm1/2$, the two-particle Fock-state expansions have the following forms~\cite{Lu:2016vqu}:
	\begin{align}
	|\Psi^{\uparrow(\downarrow)}_{\text{two\, particle}}(p^+,\bm{p}_\perp=\bm{0}_\perp) \rangle =&\int \frac{d^2\bm{k}_\perp dx}{16\pi^3\sqrt{x(1-x)}}\nonumber \\
	&\times \left[\psi^{\uparrow(\downarrow)}_{+1+\frac{1}{2}}(x,\bm{k}_\perp)
	\left|+1,+\frac{1}{2},xp^+,\bm{k}_\perp \right\rangle+\psi^{\uparrow(\downarrow)}_{+1-\frac{1}{2}}(x,\bm{k}_\perp)
	\left|+1,-\frac{1}{2},xp^+,\bm{k}_\perp \right\rangle \nonumber \right.\\		&\left.+\psi^{\uparrow(\downarrow)}_{-1+\frac{1}{2}}(x,\bm{k}_\perp)
	\left|-1,+\frac{1}{2},xp^+,\bm{k}_\perp \right\rangle+\psi^{\uparrow(\downarrow)}_{-1-\frac{1}{2}}(x,\bm{k}_\perp)
	\left|-1,-\frac{1}{2},xp^+,\bm{k}_\perp \right\rangle \right],
\end{align}
where $\psi_{s_g^zs_X^z}^{\uparrow(\downarrow)}(x,\bm{k}_\perp)$ are the LFWFs for the two-particle state $|s_g^z,s_X^z,xp^+,\bm{k}_\perp\rangle$, with $s_g^z$ and $s_X^z$ the longitudinal spin components of the gluon and spectator. Motivated by the LFWFs of the electron Fock state (spin-1 photon + spin-1/2 electron) \cite{Brodsky:2000ii}, the LFWFs for $J_z=+1/2$ (up) and $J_z=-1/2$ (down) protons are:
	\begin{align}		
	\psi^\uparrow_{+1+\frac{1}{2}}(x,\bm{k}_\perp)&=-\sqrt{2}
	\frac{-k^1_\perp+ik^2_\perp}{x(1-x)}\phi,\nonumber\\
	\psi^\uparrow_{+1-\frac{1}{2}}(x,\bm{k}_\perp)&=-\sqrt{2}
	\left(M-\frac{M_X}{1-x}\right)\phi,\nonumber\\		\psi^\uparrow_{-1+\frac{1}{2}}(x,\bm{k}_\perp)&=
	-\sqrt{2}\frac{+k^1_\perp+ik^2_\perp}{x}\phi,\nonumber\\
	\psi^\uparrow_{-1-\frac{1}{2}}(x,\bm{k}_\perp)&=0,
	\label{eq:wavefunction+}
\end{align}
and
\begin{align}
	\psi^\downarrow_{+1+\frac{1}{2}}(x,\bm{k}_\perp)&=0,\nonumber\\		\psi^\downarrow_{+1-\frac{1}{2}}(x,\bm{k}_\perp)&=
	-\sqrt{2}\frac{-k^1_\perp+ik^2_\perp}{x}\phi,\nonumber\\		\psi^\downarrow_{-1+\frac{1}{2}}(x,\bm{k}_\perp)&=
	-\sqrt{2}\left(M-\frac{M_X}{1-x}\right)\phi,\nonumber\\		\psi^\downarrow_{-1-\frac{1}{2}}(x,\bm{k}_\perp)&=
	-\sqrt{2}\frac{+k^1_\perp+ik^2_\perp}{x(1-x)}\phi,
	\label{eq:wavefunction-}
\end{align}
where $M_X$ is the spectator mass parameter satisfying $M_X>M$ for proton stability, and $\phi\equiv\phi(x,\bm{k}^2_\perp)$ represents the momentum-space wave function:
\begin{align}
	\phi(x,\bm{k}^2_\perp)=\frac{g(k^2) \sqrt{x} (x-1)}{\bm{k}_\perp^2+L_X^2(M_g^2)}.
\end{align}
The coupling $g(k^2)$ at the nucleon-gluon-spectator vertex characterizes nonperturbative color interactions; we adopt a dipolar form factor \cite{Bacchetta:2020vty,Bacchetta:2024fci}:
\begin{align}
	g(k^2)=N_g \frac{k^2}{|k^2-\Lambda_X^2|^2}=N_g\frac{k^2(1-x)^2}{(\bm{k}_\perp^2+L_X^2(\Lambda_X^2))^2},
\end{align}
with $N_g$ the normalization constant and $\Lambda_X$ the cut-off parameter. The dipolar form factor ensures the convergence of $\bm{k}_\perp$ integrals and suppresses high-$\bm{k}_\perp^2$ regions where the TMD formalism is invalid. The spectator on-shell condition $(P-k)^2=M_X^2$ implies the gluon off-shell condition:
\begin{align}
	k^2&\equiv\tau(x,\bm{k}_\perp^2)=-\frac{\bm{k}_\perp^2+L_X^2(M_g^2)}{1-x}+M_g^2,\nonumber\\
	L_X^2(M_g^2)&=xM_X^2+(1-x)M_g^2-x(1-x)M^2,
\end{align} 
and thus
\begin{align}
	L_X^2(\Lambda_X^2)=xM_X^2+(1-x)\Lambda_X^2-x(1-x)M^2.
\end{align}
We set the gluon mass $M_g=0\,\text{GeV}$ throughout this work.

Using the LFWFs of the proton in Eqs.~(\ref{eq:wavefunction+}-\ref{eq:wavefunction-}), the unpolarized and helicity TMDs of gluons in the overlap representation are expressed as
\begin{align} f_1^g(x,\bm{k}_\perp^2)=&\frac{1}{16\pi^3}\sum_{s_{g}^{z},s_X^z}\psi^{\uparrow\star}_{s_{g}^zs_X^z}(x,\bm{k}_\perp) \psi^{\uparrow}_{s_{g}^zs_X^z}(x,\bm{k}_\perp)\nonumber\\ =&\frac{N_g^2(1-x)^2}{8\pi^3x}\frac{(1+(1-x)^2)\bm{k}_\perp^2+x^2(M_X-(1-x)M)^2}{(\bm{k}_\perp^2+L_X^2(\Lambda_X^2))^4},\\
	g_{1L}^g(x,\bm{k}_\perp^2)=&\frac{1}{16\pi^3}\sum_{s_X^z}[\psi^{\uparrow\star}_{+1 s_X^z}(x,\bm{k}_\perp) \psi^{\uparrow}_{+1 s_X^z}(x,\bm{k}_\perp)-\psi^{\uparrow\star}_{-1 s_X^z}(x,\bm{k}_\perp) \psi^{\uparrow}_{-1 s_X^z}(x,\bm{k}_\perp)]\nonumber\\
	=&\frac{N_g^2(1-x)^2}{8\pi^3}\frac{(2-x)\bm{k}_\perp^2+x(M_X-(1-x)M)^2}{(\bm{k}_\perp^2+L_X^2(\Lambda_X^2))^4}.
\end{align}	
After performing the integration over $\bm{k}_\perp$, we obtain the corresponding collinear unpolarized and helicity PDFs
\begin{align}
	f_1^g(x)&=\frac{N_g^2 (1-x)^2}{48 \pi^2 x}\frac{x[(2+x^2)M_X^2-(1-x)(2-4x+3x^2)M^2-4x(1-x)M M_X]+(1-x)(2-2x+x^2)\Lambda_X^2}{[L_X^2(\Lambda_X^2)]^3},
	\label{eq:f1gx}\\
	g_{1L}^g(x)&=\frac{N_g^2 (1-x)^2}{48 \pi^2 }\frac{x[(4-x)M_X^2-x(1-x)M^2-4(1-x)M M_X]+(1-x)(2-x)\Lambda_X^2}{[L_X^2(\Lambda_X^2)]^3}.
	\label{eq:g1Lgx}
\end{align}

The gluon Sivers function can be calculated in the overlap representation according to~\cite{Lu:2016vqu,Lu:2006kt,Bacchetta:2008af}
\begin{align}
	\frac{k_\perp^1-ik_\perp^2}{2M}f_{1T}^{\perp g}(x,\bm{k}_\perp^2)=i\sum_{s_{g}^{z},s_X^z}\int\frac{d^2\bm{k}_\perp^\prime}{16\pi^3}\psi^{\uparrow\star}_{s_{g}^zs_X^z}(x,\bm{k}_\perp)G(x,\bm{k}_\perp,\bm{k}_\perp^\prime) \psi^{\downarrow}_{s_{g}^zs_X^z}(x,\bm{k}^\prime_\perp),\label{eq:siverskprime}
\end{align}	
where $G(x,\bm{k}_\perp,\bm{k}^\prime_\perp)$ is the kernel that describes the final-state interaction between the struck gluon and the spectator. This kernel has been extracted from the calculation of the gluon Sivers function in the quark target model~\cite{Meissner:2007rx,Goeke:2006ef}:
\begin{align}
	G(x,\bm{k}_\perp,\bm{k}_\perp^\prime)=\frac{-iC_A\alpha_s}{4\pi(\bm{k}_\perp-\bm{k}_\perp^\prime)^2},
\end{align}	
which is assumed to be the same as that in the light-front spectator model~\cite{Lu:2016vqu}, although the interaction kernel is typically model-dependent. After evaluating the integration over $\bm{k}_\perp^\prime$ in Eq.~(\ref{eq:siverskprime}), we obtain the analytical expression of the gluon Sivers function in this model
\begin{align}
	f_{1T}^{\perp g}(x,\bm{k}_\perp^2)=\frac{N_g^2C_A\alpha_s(1-x)^3}{2(2\pi)^3}\frac{((1-x)M-M_X)M}{L_X^2(\Lambda_X^2)(\bm{k}_\perp^2+L_X^2(\Lambda_X^2))^3}.
	\label{eq:sivers}
\end{align}	

The gluon GTMD $F^g_{1,4}$ is extracted from the off-forward correlator in Eq.~(\ref{eq:W}) via the helicity difference:
\begin{align} W_{\uparrow\uparrow}-W_{\downarrow\downarrow}=
\frac{2i\epsilon_\perp^{ij}k_\perp^i\Delta_\perp^j}{M^2\sqrt{1-\xi^2}}F_{1,4}^g.
\end{align}	
As a complex-valued function, $F^g_{1,4}=\text{Re}F^g_{1,4}+i\text{Im}F^g_{1,4}$. 
As mentioned in Ref.~\cite{Meissner:2009ww}, the imaginary part of the gluon GTMD $F^g_{1,2}$ is linked to the Sivers function in the forward limit ($\xi\to0$, $\bm{\Delta}_\perp\to0$) via
\begin{align}
	\text{Im}F^g_{1,2}(x,0,\bm{k}_\perp^2,0,0)=-f_{1T}^{\perp g}(x,\bm{k}_\perp^2),
\end{align}	
while the overlap representation of $\text{Re}F^g_{1,2}$ in the forward limit has the form
\begin{align}	-\frac{k_\perp^1-ik_\perp^2}{2M}\text{Re}F^g_{1,2}(x,0,\bm{k}_\perp^2,0,0)=\frac{1}{16\pi^3}\sum_{s_{g}^{z},s_X^z}\psi^{\uparrow\star}_{s_{g}^zs_X^z}(x,\bm{k}_\perp) \psi^{\downarrow}_{s_{g}^zs_X^z}(x,\bm{k}_\perp).
\end{align}	
Comparing this result with Eq.~(\ref{eq:siverskprime}), we assume that $\text{Im}F^g_{1,4}$ can be also obtained by inserting the interaction kernel $G(x,\bm{k}_\perp^{\text{out}},\bm{k}_\perp^{\text{in}})$ between the initial-state and final-state LFWFs of the proton in the overlap representation of $\text{Re}F^g_{1,4}$, i.e.,
\begin{align} \text{Re}W_{\lambda\lambda}&=\frac{1}{16\pi^3}\sum_{s_{g}^{z},s_X^z}\psi^{\lambda\star}_{s_{g}^zs_X^z}(x^{\text{out}},\bm{k}_\perp^{\text{out}}) \psi^{\lambda}_{s_{g}^zs_X^z}(x^{\text{in}},\bm{k}_\perp^{\text{in}}),\\
\text{Im}W_{\lambda\lambda}&=\frac{1}{16\pi^3}i\sum_{s_{g}^{z},s_X^z}\psi^{\lambda\star}_{s_{g}^zs_X^z}(x^{\text{out}},\bm{k}_\perp^{\text{out}})G(x,\bm{k}_\perp^{\text{out}},\bm{k}_\perp^{\text{in}}) \psi^{\lambda}_{s_{g}^zs_X^z}(x^{\text{in}},\bm{k}_\perp^{\text{in}}),
\end{align}
where the arguments of the initial-state and final-state wave functions are given by
\begin{align}
	\bm{k}^{\text{in}}_\perp=&\bm{k}_\perp - (1-x^{\text{in}})\frac{\bm{\Delta}_\perp}{2},\qquad \text{with} \qquad x^{\text{in}}=\frac{x+\xi}{1+\xi}, \nonumber\\
	\bm{k}^{\text{out}}_\perp=&\bm{k}_\perp +(1-x^{\text{out}}) \frac{\bm{\Delta}_\perp}{2},\qquad \text{with} \qquad x^{\text{out}}=\frac{x-\xi}{1-\xi}.
\end{align}
Thus, the analytical results for the real and imaginary parts of $F_{1,4}^g$ have the following forms:
\begin{align}
\text{Re}F^g_{1,4}&=\frac{32N_g^2(1-x)^3(1-\xi^2)^{\frac{3}{2}}(2x-x^2-\xi^2)M^2}{\pi^3\sqrt{\frac{x+\xi}{1+\xi}}\sqrt{\frac{x-\xi}{1-\xi}}D(x,\xi,\bm{k}_\perp,\bm{\Delta}_\perp)},\\
\text{Im}F^g_{1,4}&=\frac{8N_g^2C_A\alpha_s(1-x)(1-\xi^2)^{\frac{7}{2}}(2x-x^2-\xi^2)M^2}{\pi^4\bm{\Delta}_\perp^2\sqrt{\frac{x+\xi}{1+\xi}}\sqrt{\frac{x-\xi}{1-\xi}}D(x,\xi,\bm{k}_\perp,\bm{\Delta}_\perp)},
\end{align}
where
\begin{align} D(x,\xi,\bm{k}_\perp,\bm{\Delta}_\perp)=&\{4(x-\xi)(1-\xi)M_X^2-4(1-x)(x-\xi)M^2+4(1-x)(1-\xi)\Lambda_X^2
+[2(1-\xi)\bm{k}_\perp+(1-x)\bm{\Delta}_\perp]^2\}^2\nonumber\\ &\times\{4(x+\xi)(1+\xi)M_X^2-4(1-x)(x+\xi)M^2+4(1-x)(1+\xi)\Lambda_X^2
+[2(1+\xi)\bm{k}_\perp-(1-x)\bm{\Delta}_\perp]^2\}^2.
\end{align}

\section{Parameter fitting and numerical results}\label{Sec:4}

\subsection{Parameter fitting}

The model contains three free parameters: $N_g$ (normalization), $M_X$ (spectator mass), and $\Lambda_X$ (cut-off), which are fixed by fitting the unpolarized gluon PDF to experimental data, subject to the proton stability condition $M_X>M$. As the unpolarized gluon PDF is better constrained than the helicity PDF, we fit Eq.~(\ref{eq:f1gx}) to the NNPDF3.1 dataset for $xf_1^g(x)$~\cite{NNPDF:2017mvq} at the initial scale $\mu_0=2\,\text{GeV}$. We use 1000 data points in the range $0.001\leq x\leq1$ (step size 0.001). 
Simultaneous fitting of the unpolarized and helicity PDFs to NNPDF3.1 and NNPDFpol1.1~\cite{Nocera:2014gqa} was attempted but yielded significant deviations in the small-$x$ region ($0.001<x<0.01$), so we focus on the unpolarized PDF fit for parameter determination.

\begin{figure}
	\centering
	\includegraphics[width=0.45\columnwidth]{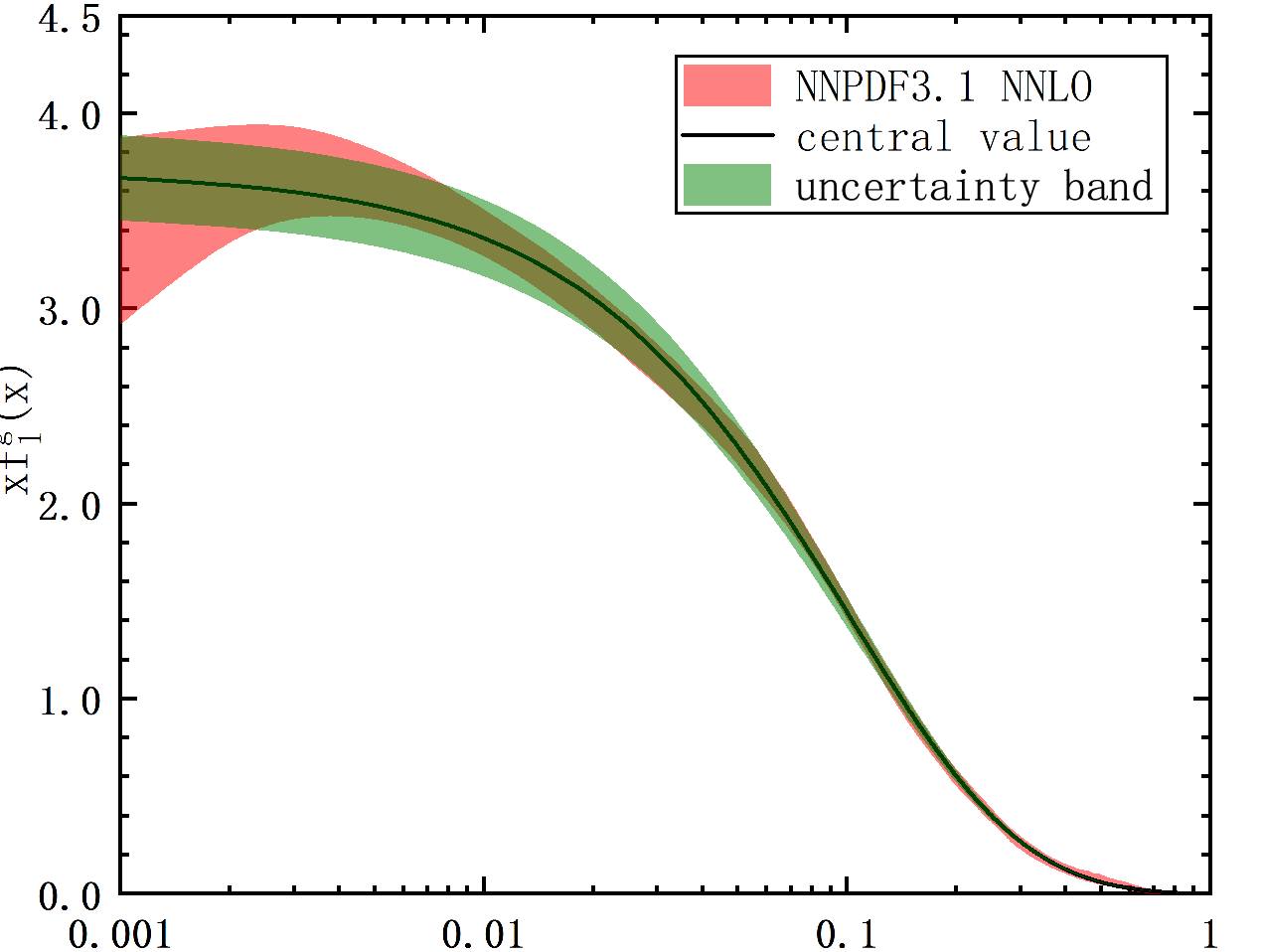}
	\caption{The fitting of the model result in Eq.~(\ref{eq:f1gx}) to the NNPDF3.1 for the unpolarized gluon PDF $f_1^g(x)$. The red band is the NNPDF3.1 uncertainty, the green band is the model's uncertainty, and the black line is the central value.}
	\label{fig:f1g}
\end{figure}

The best-fit parameter values with uncertainties are listed in Table~\ref{tab1}, and the comparison between the fitted result and the NNPDF3.1 parametrization is shown in Fig.~\ref{fig:f1g}. The red band in Fig.~\ref{fig:f1g} is the NNPDF3.1 uncertainty for $xf_1^g(x)$, the green band is our model's uncertainty, and the solid line is the central value. Despite the model's simplicity (three free parameters), it provides an fairly good description of the NNPDF3.1 dataset across the entire $x$ range, validating the choice of the light-front spectator model and dipolar form factor.

\begin{table}[htbp]
	\centering
	\caption{Best-fit values and uncertainties for the model free parameters (proton mass $M=0.938\,\text{GeV}$).}
	\label{tab1}
	\begin{tabular}{lcc} 
		\hline   
		Parameter & Central value & Uncertainty \\ 
		\hline 
		~~~~~$N_g$ & 6.717  & $\pm0.146$    \\
		~~~~~$M_X$ & 1.370  & $\pm0.010$   \\
		~~~~~$\Lambda_X$ & 0.476 & $\pm0.005$   \\
		\hline 
	\end{tabular}
\end{table}

\subsection{Numerical results for gluon distributions}

We first calculate the average longitudinal momentum fraction of gluons in the proton, defined as the second Mellin moment of the unpolarized gluon PDF:
\begin{align}
	\langle x\rangle_g=\int^1_0dx xf_1^g(x)=0.411\pm0.022.
	\label{eq:xg}
\end{align}
In Table~\ref{tab2}, we compare the numerical result in Eq.~(\ref{eq:xg}) with those obtained from other theoretical models~\cite{Chakrabarti:2023djs,Bacchetta:2020vty,Lu:2016vqu} and the lattice calculation~\cite{Alexandrou:2020sml}. 

\begin{table}[htbp]
	\centering
	\caption{Comparison of the numerical results of the average longitudinal momentum fraction of gluons.}
	\label{tab2}
	\begin{tabular}{lccccc} 
		\hline   
		& This work & \cite{Chakrabarti:2023djs} & \cite{Bacchetta:2020vty} & \cite{Lu:2016vqu} & Lattice result~\cite{Alexandrou:2020sml}\\ 
		\hline 
		$\langle x\rangle_g$ & 0.411(22) &  $0.416^{+0.048}_{-0.041}$  & 0.424(9) & 0.411 & 0.427(92)\\
		\hline 
	\end{tabular}
\end{table}

\begin{figure}
	\centering
	\includegraphics[width=0.43\columnwidth]{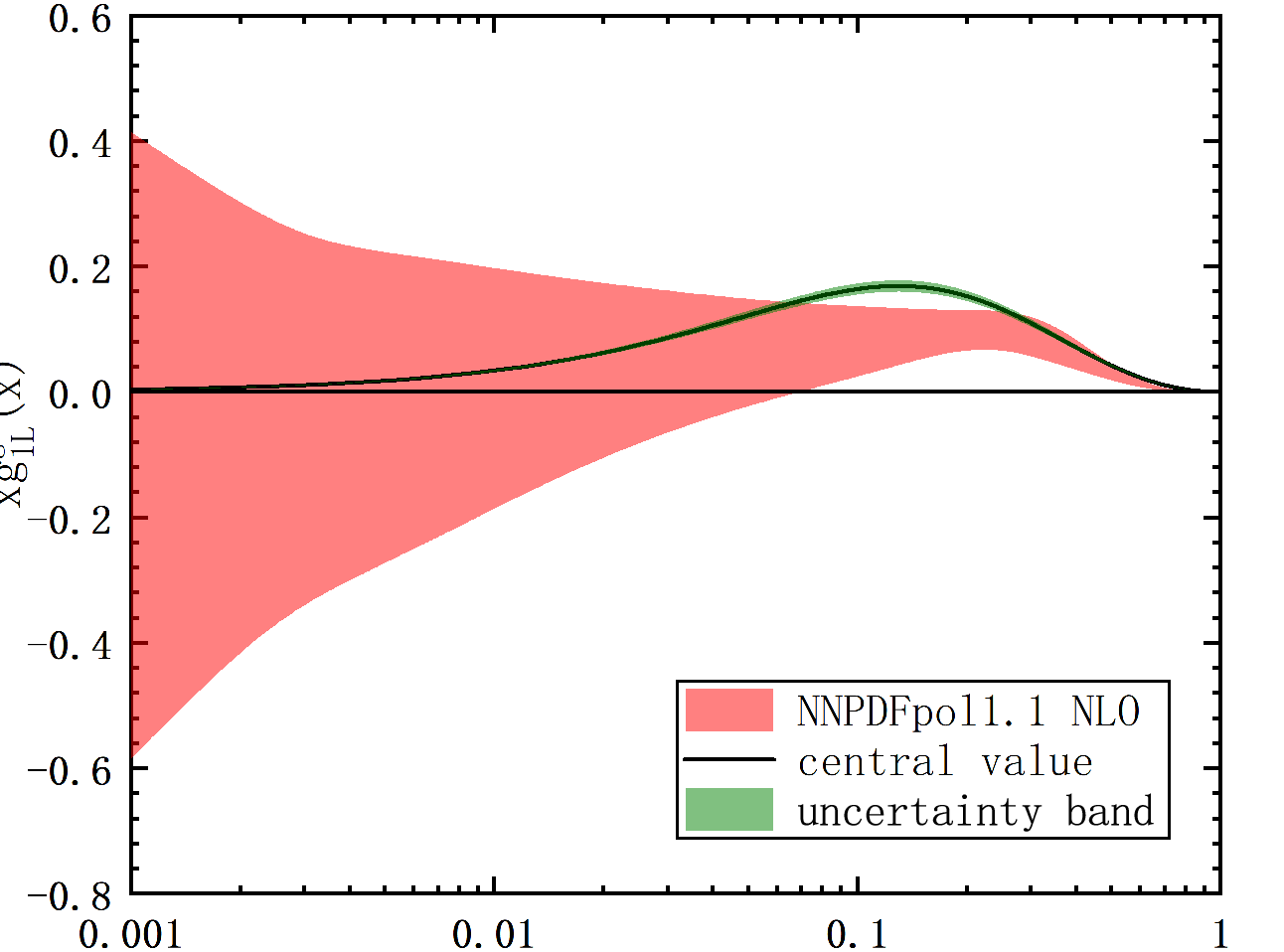}
	\includegraphics[width=0.43\columnwidth]{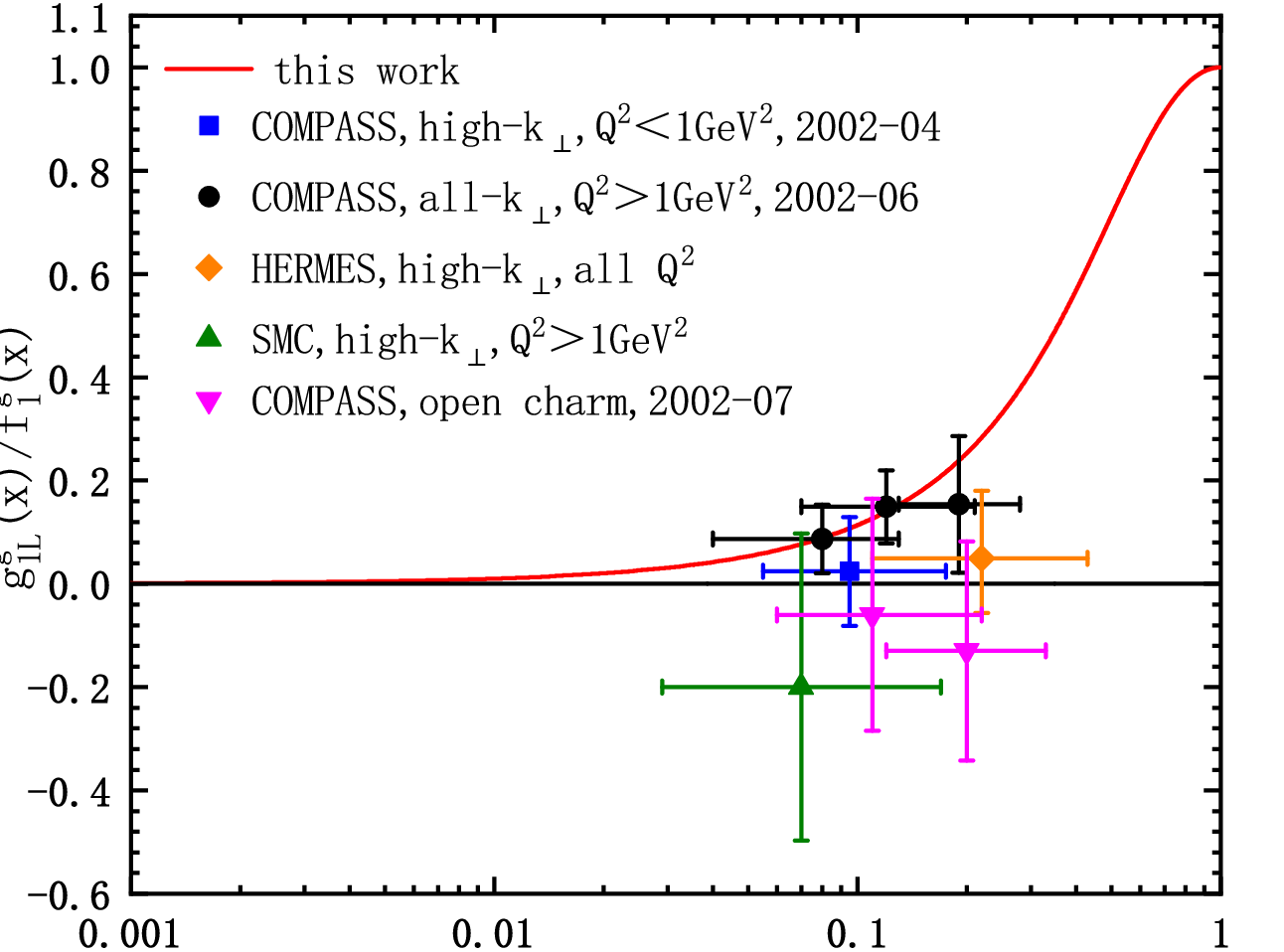}
	\caption{Left: Model prediction for the gluon helicity PDF $xg_{1L}^g(x)$ at $\mu_0=2\,\text{GeV}$, compared with the NNPDFpol1.1 dataset (red band). Green band: model uncertainty; solid line: central value. Right: Gluon helicity asymmetry ratio $g_{1L}^g(x)/f_1^g(x)$, compared with experimental data from COMPASS \cite{COMPASS:2015pim,COMPASS:2005qpp,COMPASS:2012mpe}, HERMES \cite{HERMES:2010nas}, and SMC \cite{SpinMuonSMC:2004jrx}.
}\label{fig:g1g}
\end{figure}

\begin{table}[htbp]
	\centering
	\caption{Gluon spin contribution $\Delta G=\int_{x_{\text{min}}}^{x_{\text{max}}}dx g_{1L}^g(x)$ for different $x$ ranges, compared with experimental data. All results at $\mu_0=2\,\text{GeV}$.}
	\label{tab3}
	\begin{tabular}{lcc} 
		\hline   
		Gluon spin & Experimental data & This work \\ 
		\hline 
		$\Delta G=\int^{0.3}_{0.02} dx g_{1L}^g(x)$ & 0.20(10)~\cite{PHENIX:2008swq}  & 0.353(19)    \\
		$\Delta G=\int^{0.2}_{0.05} dx g_{1L}^g(x)$ & 0.23(6)~\cite{Nocera:2014gqa} & $0.215(12)$   \\
		$\Delta G=\int^{1}_{0.05} dx g_{1L}^g(x)$ & 0.19(6)~\cite{deFlorian:2014yva} & $0.318(17)$   \\
		\hline 
	\end{tabular}
\end{table}

In Table~\ref{tab3} we compare the model predictions on the gluon spin contribution to the proton spin, integrated over several $x$ intervals, with available experimental extractions~\cite{PHENIX:2008swq,Nocera:2014gqa,deFlorian:2014yva}. Our model yields a relatively larger total gluon spin contribution, with the dominant share arising from the small-$x$ region. By contrast, the spin fraction of the proton in the large-$x$ region is mainly carried by quarks. Large gluon spin values similar to ours have previously been reported in Refs.~\cite{Kaur:2019kpe,Sufian:2020wcv}. Smaller gluon spin results have also been found in the literature~\cite{Bacchetta:2020vty,Sufian:2020wcv} when both unpolarized and helicity PDFs were fitted simultaneously, improving consistency with the dataset. A lattice calculation also reported a small value of $\Delta G$ in Ref.~\cite{Yang:2016plb}.

In Fig.~\ref{fig:g1g}, we show the model predictions of the gluon helicity PDF $g_{1L}^g(x)$ (left panel) and the gluon helicity asymmetry ratio $g_{1L}^g(x)/f_1^g(x)$ (right panel) at $\mu_0=2$ GeV. In the left panel, the red band represents the gluon helicity PDF from the NNPDFpol1.1 global analysis~\cite{Nocera:2014gqa}, which has considerable uncertainty in the entire $x$ range, especially in the small-$x$ region, the green band represents the uncertainty band of the model prediction, and the black solid line represents the result of the central values of parameters. Our result is broadly consistent with the global analysis except in the interval $0.07<x<0.24$.
 
In the right panel of Fig.~\ref{fig:g1g} we show the gluon helicity asymmetry ratio $g_{1L}^g(x)/f_1^g(x)$ together with experimental measurements~\cite{COMPASS:2015pim,COMPASS:2005qpp,HERMES:2010nas,SpinMuonSMC:2004jrx,COMPASS:2012mpe}. Central-value prediction is shown by the solid line. The ratio is predicted with negligible model uncertainty and is in good agreement with the existing data. Furthermore, our calculation satisfies the model-independent QCD constraints~\cite{Brodsky:1989db,Brodsky:1994kg}:
\begin{align}
	\lim_{x\to0}\frac{g_{1L}^g(x)}{f_1^g(x)}=0\qquad\text{and}\qquad\lim_{x\to1}\frac{g_{1L}^g(x)}{f_1^g(x)}=1.
\end{align}	

\begin{figure}
	\centering
	\includegraphics[width=0.45\columnwidth]{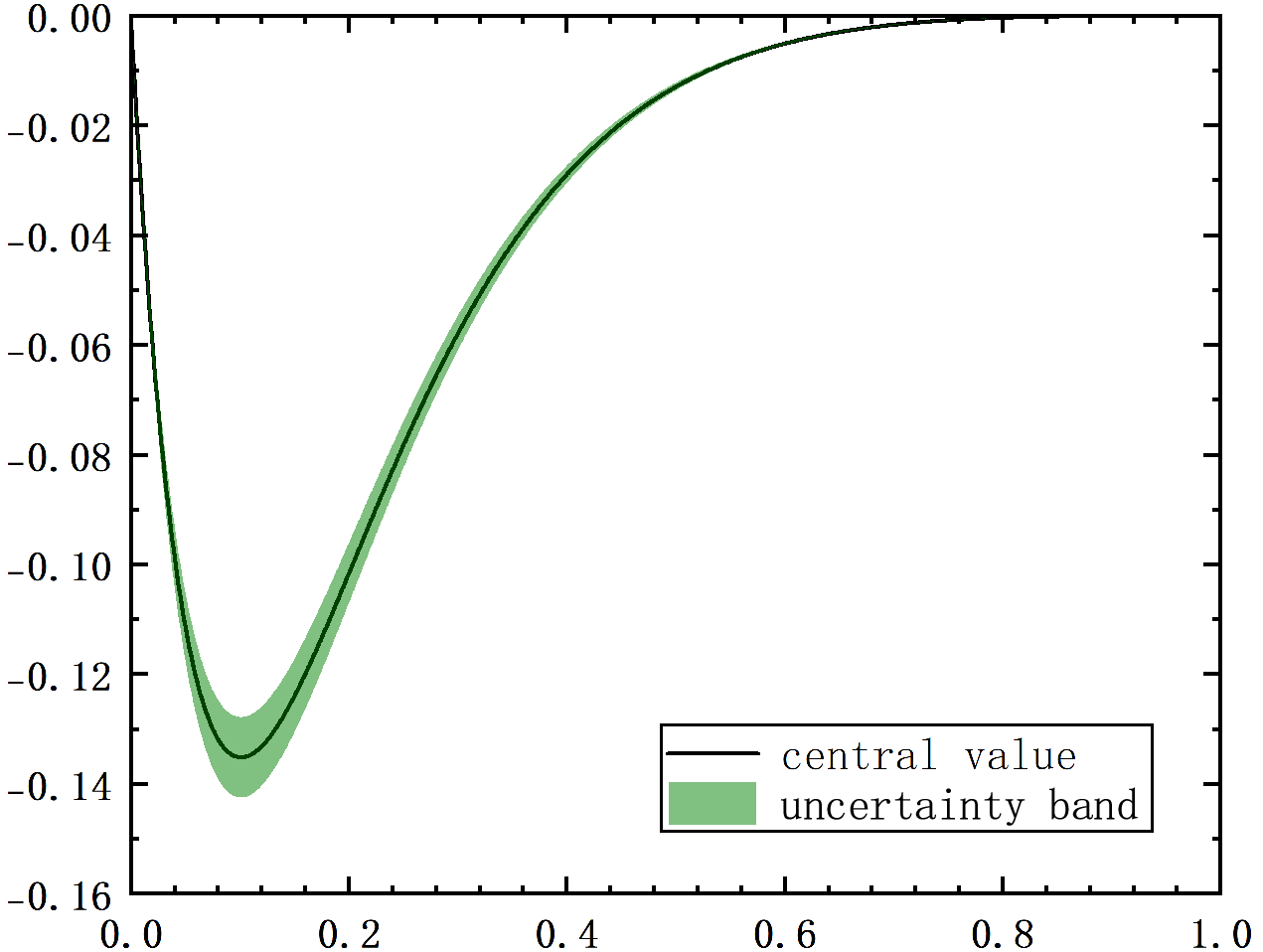}
	\caption{The dependence of the canonical gluon OAM $l_z^g(x)$ (timed with $x$) on $x$ in the proton.}
	\label{fig:xlzg}
\end{figure}

The model result for the total gluon spin is
\begin{align}
	\Delta G=\int^1_0 dx g_{1L}^g(x)=0.469\pm0.025,
	\label{eq:DeltaG}
\end{align}
while the canonical gluon  OAM is calculate numerically from our model:
\begin{align}
	l_z^g=-\int^1_0 dx\int d^2\bm{k}_\perp\frac{\bm{k}_\perp^2}{M^2}F^g_{1,4}(x,0,\bm{k}_\perp^2,0,0)=-0.370\pm0.020.
	\label{eq:lzg}
\end{align}
which is negative and substantial in size. After summing the contributions from the spin and the OAM in Eqs.~(\ref{eq:DeltaG}-\ref{eq:lzg}), it leads to a smaller  $J_z^g=l_z^g+\Delta G=0.099\pm0.005$, compared with the latest lattice calculations on the proton spin decomposition at the physical pion mass~\cite{Alexandrou:2020sml} which yields 
$J_z^g = 0.187(46)$ at the scale 2 GeV.  
In Fig.~\ref{fig:xlzg} we display the $x$-dependence of the canonical gluon OAM density $l_z^g(x)$. The OAM is negative across $x$ and is dominated by the small-$x$ region, mirroring the behavior of the gluon helicity distribution.

\begin{figure}
	\centering
	\includegraphics[width=0.43\columnwidth]{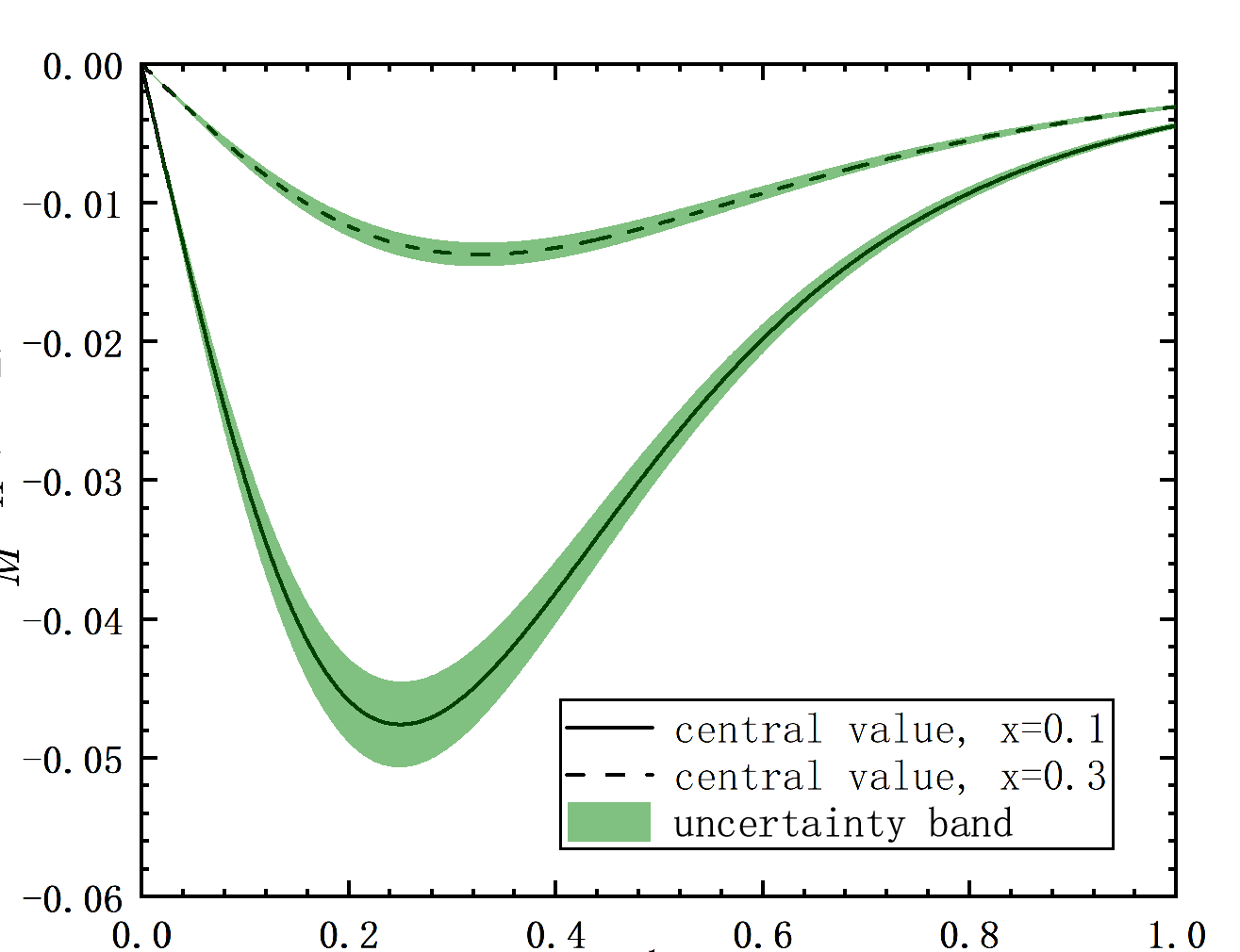}
	\includegraphics[width=0.43\columnwidth]{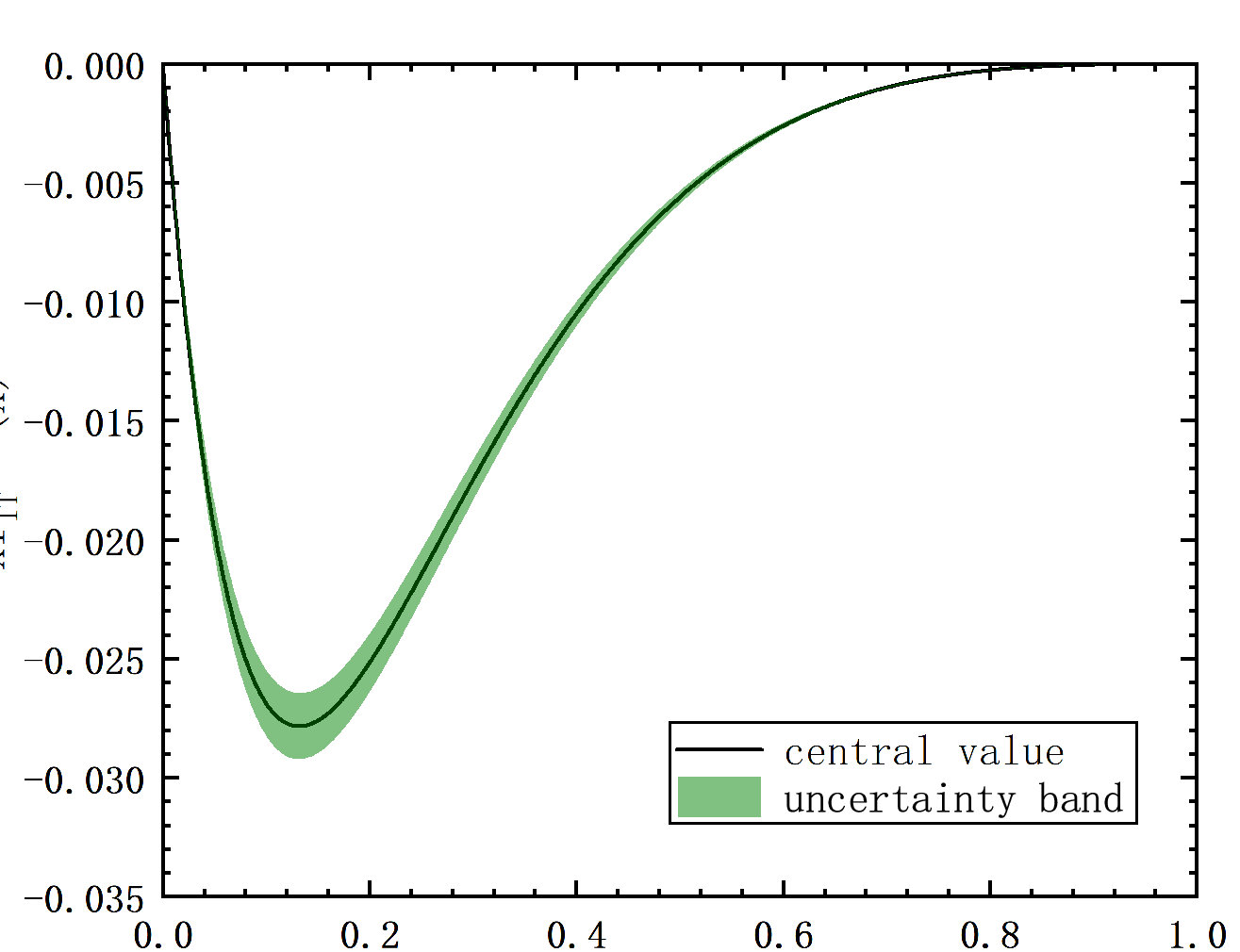}
	\caption{The dependence of the gluon Sivers function $f_{1T}^{\perp g}(x,\bm{k}_\perp^2)$ (timed with $x\frac{k_\perp}{M}$) on $k_\perp$ at $x=0.1$, $0.3$ (left panel). The dependence of the first transverse moment of the gluon Sivers function $f_{1T}^{\perp(1)g}(x)$ (timed with $x$) on $x$ (right panel).}
	\label{fig:f1Tperp}
\end{figure}

We now discuss the gluon Sivers function. Fig.~\ref{fig:f1Tperp} (left) shows $x\frac{k_\perp}{M}f_{1T}^{\perp g}(x,\bm{k}_\perp^2)$ as a function of $k_\perp$ at $x=0.1$ and $0.3$, while the right panel displays the first transverse moment of the gluon sivers function defined as
\begin{align}
	f_{1T}^{\perp(1)g}(x)=\int d^2\bm{k}_\perp\frac{\bm{k}_\perp^2}{2M^2}f_{1T}^{\perp g}(x,\bm{k}_\perp^2).
\end{align}	
From the figure, one can observe that the peak magnitude decreases and shifts to larger $k_\perp$ as $x$ increases. The Sivers function in our model is negative in the entire $x$ region, consistent with the extraction in Ref.~\cite{DAlesio:2015fwo}.

The Sivers function must satisfy the positivity bound~\cite{Bacchetta:1999kz,Mulders:2000sh},\begin{align}
	\frac{k_\perp}{M}|f_{1T}^{\perp g}(x,\bm{k}_\perp^2)|\leq f_1^g(x,\bm{k}_\perp^2).
\end{align}	
We find this inequality to hold for $k_\perp\lesssim 10\ \mathrm{GeV}$ but violated at larger $k_\perp$, a behavior similar to that observed for quark Sivers functions in quark–diquark models~\cite{Kotzinian:2008fe}. This violation likely originates from the truncation mismatch in model calculations: T-odd TMDs are computed at $\mathcal{O}(\alpha_s^0)$ while T-even TMDs are kept at $\mathcal{O}(\alpha_s^0)$~\cite{Pasquini:2011tk}. Given that in our model gluon TMDs are numerically negligible ($\leq10^{-6}$) for $k_\perp>10\ \mathrm{GeV}$ and the model average transverse momentum is small ($\langle k_\perp^g\rangle=70\pm4\ \mathrm{MeV}$), the region of validity is effectively limited to moderate $k_\perp$, and the observed partial violation at very large $k_\perp$ does not compromise the phenomenological applicability of the model within its expected domain. 

\subsection{The $\langle\sin(2\phi)\rangle$ azimuthal asymmetry}

\begin{figure}
	\centering
	\includegraphics[width=0.45\columnwidth]{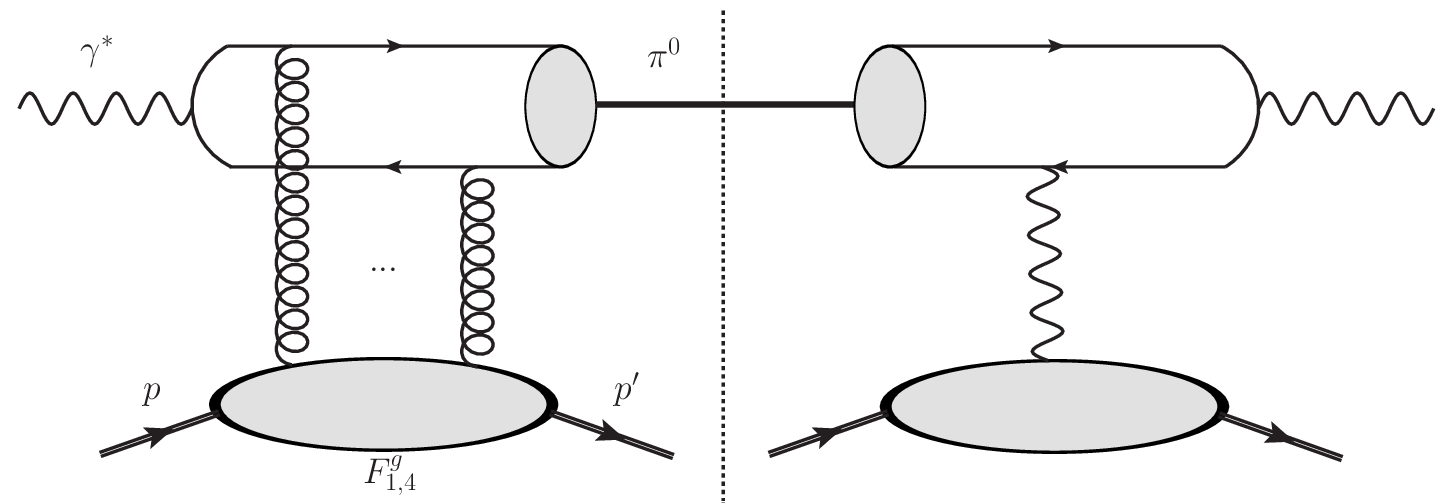}
	\caption{Coulomb-nuclear interference contribution to the exclusive $\pi^0$ production in electron-(longitudinally polarized) proton collisions.}
	\label{fig:interference}
\end{figure}

To present numerical results for the $\langle\sin(2\phi)\rangle$ azimuthal asymmetry we first compute the differential cross section for exclusive $\pi^0$ production in $ep$ collisions with unpolarized lepton and proton beams~\cite{Boussarie:2019vmk}. The unpolarized amplitude receives contributions from $t$-channel exchange of the spin-dependent Odderon, which is related to the $d$-type gluon Sivers function $f_{1T}^{\perp g(d)}(x,\bm{k}_\perp^2)$ and has been proposed as a classic channel to search for the Odderon~\cite{Berger:1999ca}. The corresponding diagram is shown in the left hand side of the cut in Fig.~\ref{fig:interference}. In the forward and perturbative-$Q^2$ limit the Odderon-induced cross section can be written as~\cite{Bhattacharya:2023yvo}
\begin{align} \frac{d\sigma^{\text{Odderon}}}{dtdQ^2dx_B}\approx\frac{\pi^5\alpha^2_{\text{em}}\alpha_s^2f_\pi^2}{8x_BN_c^2M^2Q^6}\bigg[1-y+\frac{y^2}{2}\bigg]\bigg[\int^1_0dz\frac{\phi_\pi(z)}{z(1-z)}\int d^2\bm{k}_\perp\frac{\bm{k}_\perp^2xf_{1T}^{\perp g(d)}(x,\bm{k}_\perp^2)}{\bm{k}_\perp^2+z(1-z)Q^2}\bigg]^2,
	\label{eq:odderon}
\end{align}	
where $N_c=3$, $\alpha_{\text{em}}=1/137$, and $f_\pi=131$ MeV is the $\pi^0$ decay constant. We also set $\alpha_s=0.3$. For the distribution amplitude (DA) $\phi_\pi(z)$ of the pion, we adopt the asymptotic form $6z(1-z)$~\cite{Bhattacharya:2023hbq}. The expression in Eq.~(\ref{eq:odderon}) is valid in the forward region $t\approx0$ when $Q^2$ is perturbative. 
Given that Eq.~(\ref{eq:odderon}) is obtained in the eikonal approximation $t\approx-\bm{\Delta}_\perp^2$, the variables $x$ and $\xi$ follow the specific approximations: $x\approx x_B/2$ and $\xi\approx x$~\cite{Boussarie:2019vmk}.

The exclusive $\pi^0$ production in $ep$ collisions also receives contributions from the single-photon mediated Primakoff process~\cite{Primakoff:1951iae,Liping:2014wbp,Kaskulov:2011ab,Lepage:1980fj,
Khodjamirian:1997tk,Jia:2022oyl} in the region $\Delta_T$ is very small, as shown in the right hand side of the cut in Fig.~\ref{fig:interference}. The Primakoff contribution in terms of the pion DA reads
\begin{align} \frac{d\sigma^{\text{Pri}}}{dtdQ^2dx_B}\approx\frac{\alpha^4_{\text{em}}(2\pi)[1+(1-y)^2]f_\pi^2}{x_BQ^6\bm{\Delta}_\perp^2}\frac{1-\xi}{1+\xi}\mathcal{F}^2(t)\bigg[\int^1_0\frac{dz}{6z(1-z)}\phi_\pi(z)\bigg]^2,
\end{align}	
where $\mathcal{F}(t)=1/(1+\frac{-t}{Q^2_0})^2$ with $Q^2_0=0.71\,\text{GeV}^2$. It is evident that this process becomes increasingly dominant as the transverse momentum transfer decreases due to the $\bm{\Delta}_\perp^2$ term in the denominator. As noted in Ref.~\cite{Boussarie:2019vmk}, there is no interference between the spin-dependent Odderon and the Primakoff amplitude as the proton helicity flips in the former but not in the latter. 

For a longitudinally polarized proton target, the interference amplitudes (Fig.~\ref{fig:interference}) between
two gluon exchange diagram and the Primakoff process gives rise to the $\sin2\phi$ asymmetry in the exclusive production of $\pi^0$~\cite{Bhattacharya:2023yvo}. The differential cross section of the Coulomb-nuclear interference effect can be expressed in terms of $F_{1,4}^g$ as
\begin{align}
	\frac{d\Delta\sigma}{dtdQ^2dx_Bd\phi}=-\sin(2\phi)\frac{\alpha^3_{\text{em}}\alpha_s f_\pi^2(1-y)\xi x_B\mathcal{F}(t)}{3Q^8N_c}\bigg[\int^1_0dz\frac{\phi_\pi(z)}{z(1-z)}\bigg]^2\text{Im}\bigg[\int_{-1}^1dx\frac{F_{1,4}^{(1)g}(x,\xi,\bm{\Delta}_\perp)/M^2}{(x+\xi-i\epsilon)^2(x-\xi+i\epsilon)^2}\bigg],
	\label{eq:csF}
\end{align}	
where 
\begin{align}
 F^{(1)g}_{1,4}(x,\xi,\bm{\Delta}_\perp)=\int d^2\bm{k}_\perp\bm{k}_\perp^2F^g_{1,4}(x,\xi,\bm{k}_\perp^2,\bm{\Delta}_\perp^2,\bm{k}_\perp\cdot\bm{\Delta}_\perp),
\end{align}
and $\phi$ is the azimuthal angle between the transverse momenta of the scattered electron $\bm{l}^\prime_\perp$ and the recoil proton $\bm{\Delta}_\perp$. Similar $\sin(2\phi)$ azimuthal angular correlation also appears in the SIDIS process, providing a probe for the specific gluon TMDs~\cite{Li:2021mmi}.

Because gluons are charge-conjugation even, the $x$-integral of the real part of $F_{1,4}^{(1)g}$ in Eq.~(\ref{eq:csF}) vanishes, while that of the imaginary part survives~\cite{Bhattacharya:2023yvo}. Similar to the case of the Sivers function, the imaginary part of the GTMD $F^g_{1,4}$ is generated by an additional gluon exchange encoded in the NLO expansion of the gauge link and is related to a tri-gluon correlation. Thus $\text{Im}F_{1,4}^{(1)g}$ can be accessed via the cross section in Eq.~(\ref{eq:csF}). The coupling between the singularities of $F_{1,4}^g$ at $x=\pm\xi$ and the double poles at $x=\pm\xi$ induces the divergences of the $x$-integral, which can be regularized by introducing the mean-squared transverse momentum of the gluon inside the pion ($\langle \bm{p}_\perp^2\rangle=0.04\,\text{GeV}^2$~\cite{Bhattacharya:2023hbq,Goloskokov:2007nt,Goloskokov:2005sd}) to shift the double pole from $1/(x-\xi+ i\epsilon)^2$ to $1/(x-\xi-\langle \bm{p}_\perp^2\rangle/Q^2+ i\epsilon)^2$ (and similar for $1/(x+\xi- i\epsilon)^2$)~\cite{Anikin:2002wg}. In addition, our model is limited to the minimum Fock state of the proton (composed of three valence quarks and a gluon), so we only need to perform the $x$-integration over the region $\xi<x\leq1$.

The asymmetry is defined as the average value of the $\sin(2\phi)$ azimuthal angular correlation:
\begin{align}
	\langle \sin(2\phi)\rangle=\frac{\int\frac{d\Delta\sigma}{d\mathcal{P}.\mathcal{S}.}\sin(2\phi)d\mathcal{P}.\mathcal{S}.}{\int[\frac{d\sigma^{\text{Pri}}}{d\mathcal{P}.\mathcal{S}.}+\frac{d\sigma^{\text{Odderon}}}{d\mathcal{P}.\mathcal{S}.}]d\mathcal{P}.\mathcal{S}.}.
\end{align}
where $d\mathcal{P.S.}$ denotes the appropriate phase-space elements. Our calculation provides the first model estimate of the imaginary part of $F_{1,4}^{(1)g}$; numerically we find $\text{Im}F_{1,4}^{(1)g}<\text{Re}F_{1,4}^{(1)g}$, consistent with general expectations~\cite{Bhattacharya:2023yvo}.

\begin{figure}
	\centering
	\includegraphics[width=0.43\columnwidth]{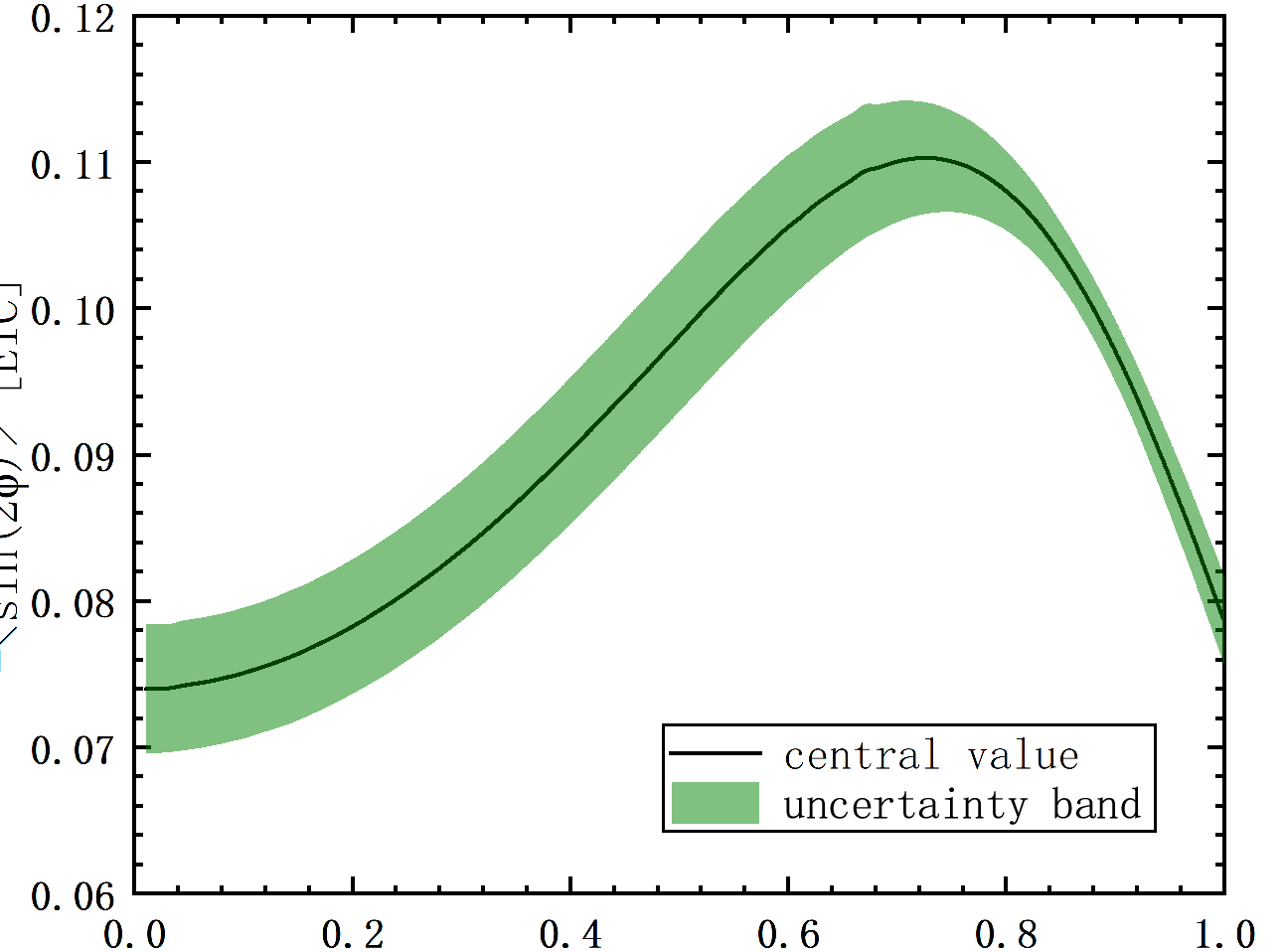}
	\includegraphics[width=0.43\columnwidth]{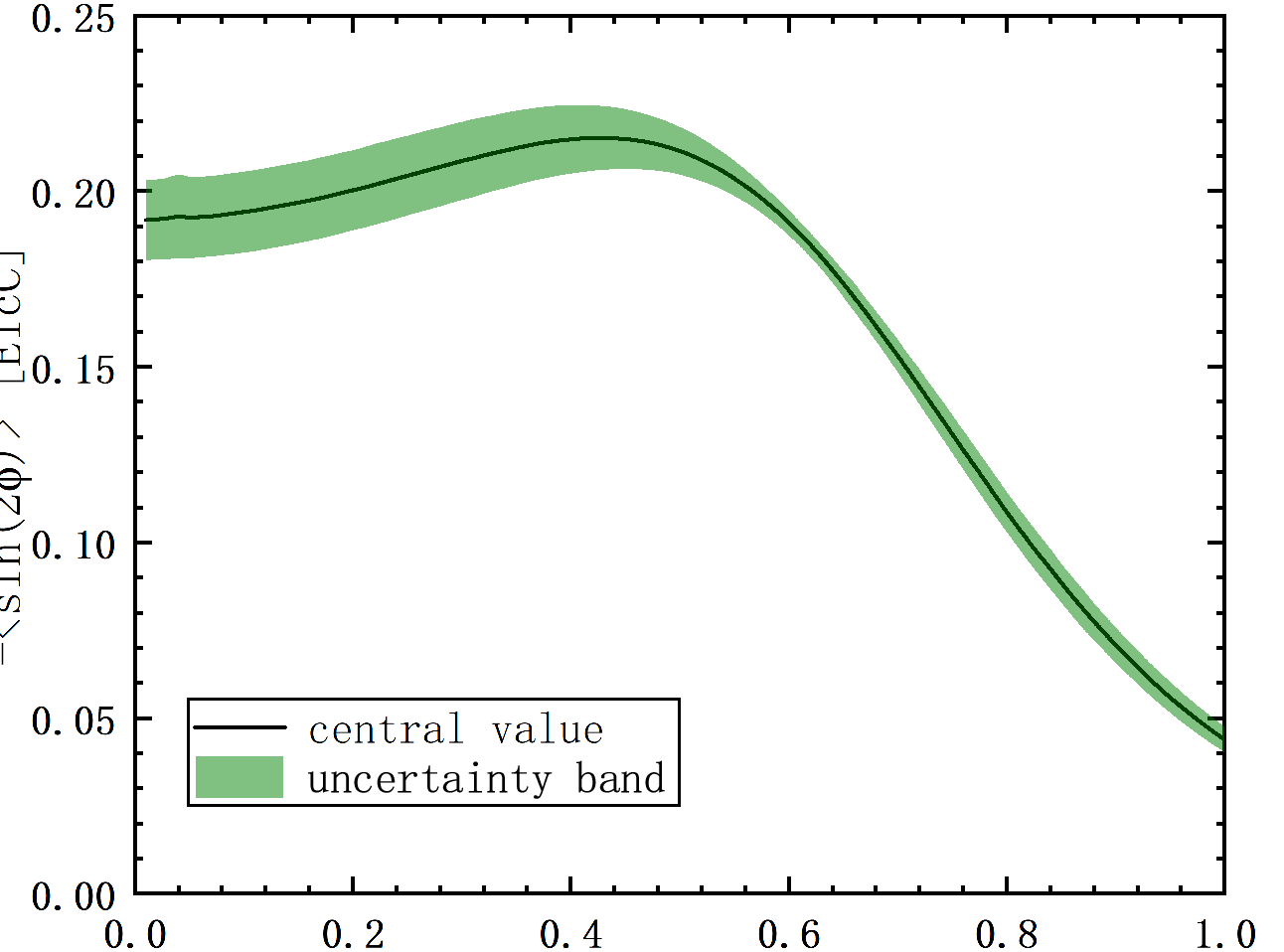}
	\caption{The azimuthal asymmetries $\langle\sin(2\phi)\rangle$ as a function of $\Delta_\perp$ for the EIC (left panel) and EicC (right panel) kinematics after integrating over $Q^2$.}
	\label{fig:sin}
\end{figure}

In Fig.~\ref{fig:sin}, we depict the $\Delta_\perp$-dependence of the predicted  $\sin(2\phi)$   azimuthal asymmetries for the EIC kinematics 
$$\sqrt{s_{ep}}=100\,\text{GeV},\,y=0.05, 1\leq Q^2\leq 15\,\text{GeV}^2$$ 
and for EicC kinematics
$$\sqrt{s_{ep}}=16\,\text{GeV},\,y=0.7, 1\leq Q^2\leq 10\,\text{GeV}^2$$  
The bands represent the uncertainty of model predictions, and the solid lines represent the results from the central values of parameters. The choices of $y$ and $Q^2$ integration limits are guided by two considerations: numerical stability of the asymmetry with respect to the upper $Q^2$ limit, and ensuring the process remains in the forward region for the entire $Q^2$ integration interval. We observe that the asymmetry peaks near $\Delta_\perp=0.7$ GeV for the EIC kinematics and near $\Delta_\perp=0.4$ GeV for the EicC kinematics. The asymmetry rapidly approaches zero with increasing $\Delta_\perp$, indicating that the dominant signal is concentrated at small transverse momentum transfer.  Additionally, we find that the magnitude of the asymmetry grows with increasing  $\xi$  (equivalently decreasing $y$), so experimental constraints on $F_{1,4}^{g}$ are more favorable at relatively large $\xi$. Note that $t\approx 0$ is a prerequisite for all discussions.
 
\section{Summary}\label{Sec:5}

In this work, we studied the azimuthal asymmetry in the exclusive $\pi^0$ production in electron-(longitudinally polarized) proton collisions, which provides access to the imaginary part of the gluon GTMD $F_{1,4}^g$ related to the canonical OAM. The asymmetry arises from the interference of the double-gluon exchange process and the Primakoff process, with its dominant contribution concentrated in the forward region. 
The double-gluon exchange process is governed by the spin-independent Odderon (or the dipole-type gluon Sivers function).

Within a light-front spectator model of the proton with gluonic degrees of freedom, we numerically evaluated the average longitudinal momentum fraction, the helicity distribution, the canonical OAM, and the Sivers function of gluon. The nucleon-gluon-spectator vertex is described by a dipolar form factor, and the model parameters are determined by fitting the unpolarized gluon PDF to the NNPDF3.1 parametrization at the initial scale $\mu_0=2\,\text{GeV}$.

The same spectator model, with the parameters constrained by the fit, yielded the estimate of the imaginary part of the GTMD $F_{1,4}^g$ and provides predictions for the resulting $\langle\sin(2\phi)\rangle$ azimuthal asymmetry in EIC and EicC kinematics. 
Our results show that the asymmetry is concentrated at small transverse momentum transfer and increases with $\xi$, suggesting favourable conditions for experimental extraction in the forward region. 
This indicates that exclusive $\pi^0$ production in $ep$ collision in the forward region is a promising channel to constrain $\text{Im}\,F_{1,4}^g$— and hence the gluon OAM — with improved prospects at relatively large $\xi$.
While model-dependent, these calculations offer useful guidance for future experimental studies and motivate further theoretical work, including the incorporation of higher-Fock contributions.

\section*{Acknowledgements}
This work is partially supported by the National Natural Science Foundation of China under grant number 12447136 and 12150013.

\end{document}